\documentclass[fleqn,usenatbib]{mnras}

\usepackage{newtxtext,newtxmath}
\usepackage[T1]{fontenc}

\DeclareRobustCommand{\VAN}[3]{#2}
\let\VANthebibliography\thebibliography
\def\thebibliography{\DeclareRobustCommand{\VAN}[3]{##3}\VANthebibliography}

\usepackage{graphicx}	% Including figure files
\usepackage{amsmath}	% Advanced maths commands

\newcommand{\colibre}{\textsc{colibre}}
\title[MQGs and super-Eddington accretion at high $z$]{The importance of super-Eddington black hole accretion for the emergence of massive quiescent galaxies at high redshift}

\author[E. Chaikin et al.]{Evgenii Chaikin,$^{1,2}$\thanks{E-mail: evgenii.chaikin@durham.ac.uk} 
Joop Schaye,$^{2}$ 
Filip Hu\v{s}ko,$^{2}$ 
Cedric G. Lacey,$^{1}$
Sylvia Ploeckinger,$^{3}$ and \newauthor
Matthieu Schaller$^{4,2}$ \\
$^{1}$Institute for Computational Cosmology, Department of Physics, University of Durham, South Road, Durham, DH1 3LE, UK \\ 
$^{2}$Leiden Observatory, Leiden University, PO Box 9513, 2300 RA Leiden, the Netherlands \\
$^{3}$Department of Astrophysics, University of Vienna, T\"{u}rkenschanzstrasse 17, A-1180 Vienna, Austria \\
$^{4}$Lorentz Institute for Theoretical Physics, Leiden University, PO Box 9506, 2300 RA Leiden, the Netherlands
}

\date{Accepted XXX. Received YYY; in original form ZZZ}

\pubyear{\the\year{}}

\begin{document}
\label{firstpage}
\pagerange{\pageref{firstpage}--\pageref{lastpage}}
\maketitle

% Abstract of the paper
\begin{abstract}
Recent \textit{JWST} observations indicate that massive quiescent galaxies (stellar mass $M_{*}\gtrsim 10^{10}~\mathrm{M_\odot}$) at high redshift ($z\gtrsim 6$) are more abundant than predicted by most existing galaxy formation simulations and semi-analytic models. Notably, the new \colibre{} simulations have succeeded in reconciling this tension, though the precise reason for their improved agreement with \textit{JWST} data remains unclear. We demonstrate that the improved agreement is largely due to super-Eddington growth of supermassive black holes (BHs) at high redshift. We run a series of $(100~\mathrm{cMpc})^{3}$ simulations with the \colibre{} subgrid physics at m7 \colibre{} resolution (gas and dark matter particle masses $m_{\rm gas}\approx m_{\rm dm}\sim 10^7~\mathrm{M_\odot}$), varying the maximum allowed BH accretion rate in units of the Eddington rate. We show that only the fiducial \colibre{} model, which permits super-Eddington accretion, is consistent with the \textit{JWST} constraints at $z \gtrsim 6$. Moreover, we find that in \colibre{} about $50$~per cent of BH mass growth at high redshift occurs in the super-Eddington regime, even though such events are extremely rare in time. Our work highlights the important role of super-Eddington accretion in simulations of galaxy formation for reproducing the observed early emergence of quenching of massive galaxies.
\end{abstract}


\section{Introduction}
\label{Section: introduction}

Massive quiescent galaxies (MQGs) detected at high redshift ($z \gtrsim 6$) provide a powerful test for theoretical models of galaxy formation. Defined as galaxies with stellar mass $M_* \gtrsim 10^{10}~\mathrm{M_\odot}$ and negligible specific star formation rate (sSFR; e.g. $\text{sSFR} < 0.2/t_{\rm age}(z)$ where $t_{\rm age}(z)$ is the age of the Universe at their observed redshift), the presence of MQGs demands that realistic galaxy formation models must be able to predict their observed abundance.

The comoving number density of MQGs, $n_{\rm MQG}$, is the fundamental statistic for comparing theoretical models with observations of MQGs. Unlike more traditional constraints imposed on theoretical models at low redshift, such as the observed galaxy stellar mass function (GSMF) and the size -- stellar mass relation, which ensure a realistic \textit{overall} galaxy population \citep[see e.g.][]{2015MNRAS.450.1937C}, the abundance of MQGs at high redshift tests galaxy formation physics under \textit{extreme} conditions. Reproducing the observed $n_{\rm MQG}$ at $z\gtrsim 6$ requires models to not only form massive galaxies within $\approx 1$~Gyr of the Big Bang but also to quench them on an even shorter time-scale.

Before the launch of \textit{JWST}, observations of MQGs were largely limited to $z \lesssim 4.5$ \citep[e.g.][]{2013ApJ...777...18M, 2017Natur.544...71G,2018A&A...618A..85S}. Those earlier data provided the first hints of a tension between theoretical galaxy formation models and observations, with the models predicting a lower $n_{\rm MQG}$ than observed at high redshift \citep[e.g.][]{2019MNRAS.490.3309M, 2020ApJ...889...93V}. The first \textit{JWST} results have exacerbated this tension, indicating that massive galaxies in the high-$z$ Universe grew and became quiescent more rapidly than previously thought \citep[e.g.][]{2023A&A...677A..88B, 2024MNRAS.534..325C, 2024Natur.628..277G,2025NatAs...9..280D}.  Compared to previous instruments such as \textit{Spitzer} and \textit{HST}/WFC3, \textit{JWST}'s improved infrared capabilities have enabled the detection of MQGs out to $z \approx 7.5$ \citep{2025ApJ...983...11W, 2025A&A...702A.270B,2026ApJ..1000L..42Y}, thereby putting even tighter constraints on theoretical models. In fact, some tentative signatures of quenching have been observed even out to $z\approx 11$ for intermediate-mass galaxies \citep[$M_*\approx 10^{9}~\mathrm{M_\odot}$;][]{2026arXiv260121833H}. The majority of existing galaxy formation simulations and semi-analytic models struggle to reproduce these new data, typically predicting $n_{\rm MQG}$ values that are $1-3$~dex lower \citep[e.g.][]{2023ApJ...947...20V,2025ApJ...983...11W,2025MNRAS.539..557B,2026MNRAS.545f2087S,2026ApJ...997..252Z,2026ApJ..1000L..42Y}, though there has been recent progress in bringing some semi-analytic models closer to the data \citep{2024MNRAS.531.3551L}.

Of particular interest is the discovery of RUBIES-UDS-QG-z7 by \citet{2025ApJ...983...11W}, an MQG at $z = 7.29$ identified within the RUBIES survey \citep{2025A&A...697A.189D} using \textit{JWST}/NIRCam and MIRI photometry, followed up with \textit{JWST}/NIRSpec PRISM spectroscopic observations. The authors estimated the RUBIES-UDS-QG-z7's stellar mass $M_* \approx 10^{10.2}~\mathrm{M_\odot}$ and sSFR $< 10^{-10}~\mathrm{yr}^{-1}$, which implies an MQG number density of $n_{\rm MQG}\sim 10^{-6}~\mathrm{cMpc}^{-3}$ at $z \sim 7$, providing the highest-redshift constraint on theoretical models to date. \citet{2025ApJ...983...11W} showed that state-of-the-art cosmological simulations that produce realistic $z \approx 0$ galaxy populations, including \textsc{eagle} \citep{2015MNRAS.446..521S} and \textsc{IllustrisTNG} \citep{2018MNRAS.473.4077P}, predict an insufficient abundance of MQGs to match their data.

Remarkably, the new \colibre{} simulations of galaxy formation \citep{2026COLIBREproject,2026COLIBREcalibration} have been able to reconcile the discrepancies with \textit{JWST} data on MQGs at high $z$ \citep{chaikin2026evolutiongalaxystellarmass, 2025arXiv251216208C}, including the tension with \citet{2025ApJ...983...11W}. However, the reason for this improved agreement compared to previous models remains to be fully understood. Several explanations have been proposed by \citet{chaikin2026evolutiongalaxystellarmass}, among which is super-Eddington gas accretion onto black holes (BHs).

Unlike many earlier galaxy simulations -- including \textsc{Illustris} \citep{2014MNRAS.444.1518V}, \textsc{HorizonAGN} \citep{2014MNRAS.444.1453D}, \textsc{eagle}, \textsc{IllustrisTNG},  \textsc{Fable} \citep{2018MNRAS.479.5385H}, and \textsc{NewHorizon} \citep{2021A&A...651A.109D} -- in \colibre{} BH accretion rates are allowed to exceed the Eddington rate, which is the rate at which radiation pressure from electron scattering balances BH gravity. Physically, this limit assumes spherical accretion, but non-spherical flows in the form of accretion-disc solutions for gas with non-zero angular momentum \citep[e.g.][]{1988ApJ...332..646A,2009ApJS..183..171S} can feature super-Eddington rates. Additionally, at sufficiently high accretion rates, photons can become trapped within the inflowing gas, reducing outward radiation pressure and allowing accretion to surpass the Eddington rate by large factors \citep[e.g.][]{1979MNRAS.187..237B,2016MNRAS.459.3738I}. The presence of super-Eddington accretion is supported by observations of high-$z$ galaxies and quasars, through inferred Eddington fractions \citep[e.g.][]{2022ApJ...941..106F,2023ApJ...959...39H,2025NatAs...9..271S} and the necessity to grow BHs within the available cosmic time to the very high inferred BH masses (up to $\sim 10^{9}~\mathrm{M_\odot}$) \cite[e.g.][]{2005ApJ...633..624V,2021ApJ...907L...1W,2022ApJ...941..106F,2024A&A...691A.145M,2024Natur.636..594J}. These results are corroborated by zoom-in hydrodynamical simulations of protoclusters \citep[e.g.][]{2024A&A...686A.256L,2024MNRAS.527.1033B,2025MNRAS.537.2559H} and by semi-analytic models \citep[e.g.][]{2016MNRAS.458.3047P,2024arXiv241214248T}.

Furthermore, those earlier simulations of galaxy formation were typically run in relatively small volumes ($\sim 100^3~\mathrm{cMpc}^3$), limiting the minimum non-zero $n_{\rm MQG}$ they could predict to $\sim 10^{-6}~\mathrm{cMpc}^{-3}$ and making their predicted $n_{\rm MQG}$ prone to cosmic variance \citep[e.g.][]{2023MNRAS.524...43T}, which could exacerbate the tension with \textit{JWST} data found in earlier comparisons. Indeed, \citet{2023MNRAS.525.5520L} used the \textsc{Flares} simulations \citep{2021MNRAS.500.2127L} -- zoom-in simulations with a parent volume of $(3.2~\mathrm{cGpc})^3$ using the original \textsc{eagle} model -- to extend the predictions of the \textsc{eagle} simulations for non-zero $n_{\rm MQG}$ from $z \approx 5$ up to $z \approx 8$. \citet{2026MNRAS.546ag214T} performed mock observations of \textsc{Flares} galaxies, finding that this larger volume does not resolve the tension with \citet{2025ApJ...983...11W}. They concluded that, while accounting for systematic uncertainties in the data may alleviate the tension, achieving full agreement likely requires modifications to the modelling of BHs and AGN feedback in \textsc{eagle}, such as allowing for super-Eddington accretion.

In this work, we investigate the importance of super-Eddington BH growth for resolving the tension between theoretical predictions and the observed $z\gtrsim 6$ abundance of MQGs. We do so by comparing predictions from the new \colibre{} simulations of galaxy formation\footnote{\href{https://colibre-simulations.org/}{https://colibre-simulations.org}} \citep{2026COLIBREproject,2026COLIBREcalibration}, using models with different maximum allowed BH accretion rates. Section \ref{section: methods} describes our methodology, Section \ref{section: results} presents the results, and Section \ref{section: conclusions} summarizes our conclusions.

\section{Methods}
\label{section: methods}

\subsection{The COLIBRE simulations}
\label{subsection: colibre}

The \colibre{} simulations are described extensively in \citet{2026COLIBREproject}. Below, we only provide a summary.

The \colibre{} simulations were run using the astrophysical code \textsc{Swift} \citep{2024MNRAS.530.2378S}, with gas hydrodynamics solved using the smoothed particle hydrodynamics (SPH) scheme \textsc{Sphenix} \citep{2022MNRAS.511.2367B}. The initial conditions were generated at $z=63$ by the \textsc{monofonIC} code \citep{2020ascl.soft08024H,2021MNRAS.500..663M} using second-order Lagrangian perturbation theory. The assumed cosmology is the $\Lambda$CDM `3x2pt + all external constraints' cosmology \citep{2022PhRvD.105b3520A}: $\Omega_{\rm m,0} = 0.306$, $\Omega_{\rm b, 0} = 0.0486$, $\sigma_8 = 0.807$, $h = 0.681$, $n_{s} = 0.967$, with a single massive neutrino species of $0.06$~eV. The \colibre{} simulations include three resolutions: m7 (gas and dark matter (DM) particle masses of $m_{\rm gas} = 1.47 \times 10^7~\mathrm{M_\odot}$ and $m_{\rm dm} = 1.94 \times 10^7~\mathrm{M_\odot}$, respectively), m6 ($m_{\rm gas} = 1.8 \times 10^6~\mathrm{M_\odot}$, $m_{\rm dm} = 2.4 \times 10^6~\mathrm{M_\odot}$), and m5 ($m_{\rm gas} = 2.3 \times 10^5~\mathrm{M_\odot}$, $m_{\rm dm} = 3.0 \times 10^5~\mathrm{M_\odot}$), with simulations at lower resolutions available in larger volumes (up to $400^3$ cMpc$^3$ for m7; see table 2 in \citealt{2026COLIBREproject}). At m7 resolution, the Plummer-equivalent gravitational softening length for baryons and DM is set to the minimum of 1.4 proper kpc (pkpc) and 3.6 comoving kpc, while at m6 (m5) resolution, it is reduced by a factor of 2 (4) relative to m7.

The radiative cooling and heating rates for primordial elements and their free electrons are calculated using the non-equilibrium thermochemistry solver \textsc{chimes} \citep{2014MNRAS.440.3349R,2014MNRAS.442.2780R}. The rates for metals are provided by \textsc{hybrid-chimes} \citep{2025MNRAS.543..891P},  which uses tabulated species fractions computed by \textsc{chimes} assuming ionization equilibrium and steady-state chemistry, corrected for the non-equilibrium electron density from hydrogen and helium. The gas chemical composition is tracked by modelling the abundances of 12 individual elements, which are diffused among SPH gas neighbours using a velocity shear-based model for turbulent mixing \citep{2026MNRAS.548ag645C}. \colibre{} models the formation and evolution of dust grains \citep{2026MNRAS.545f2040T}, which are coupled to the \textsc{chimes} solver.

Star-forming gas particles are identified using a gravitational instability criterion, following \citet{2024MNRAS.532.3299N}. The \citet{1959ApJ...129..243S} law, with a star formation efficiency per free-fall time  $\varepsilon = 0.01$, is used to compute SFRs of star-forming particles, which are stochastically converted into stars. Newly formed stellar particles represent simple stellar populations with a \citet{2003PASP..115..763C} initial stellar mass function, injecting energy, momentum, and metals into their nearby gas. Feedback from core-collapse SNe is modelled following \citet{2012MNRAS.426..140D,2023MNRAS.523.3709C}, with modifications as described in \citet{2026COLIBREproject}. \colibre{} also includes SN type-Ia feedback and three early stellar feedback processes: H~\textsc{ii} regions, stellar winds, and radiation pressure \citep{2026MNRAS.546ag268B}.

BHs are represented by collisionless BH particles. BH particles are seeded by converting the densest gas particle in a Friends-of-Friends (FoF) halo into a BH particle. At m7 resolution, seeding occurs when the halo FoF mass exceeds a threshold of $5 \times 10^{10}~\mathrm{M_\odot}$, provided the FoF halo does not already contain a BH. BH particles grow in mass by accreting surrounding gas and by merging with other BHs. The accretion rate onto BHs is computed using the modified Bondi-Hoyle-Lyttleton formula \citep{1952MNRAS.112..195B,1939PCPS...35..405H}, with turbulence and vorticity corrections from \citet{Krumholz_et_al_2006}, and is capped at 100 times the Eddington rate\footnote{The choice to allow super-Eddington BH accretion in \colibre{} is motivated by the growing observational evidence (see Section~\ref{Section: introduction}), while the cap of $10^{2}$ (which is hardly ever reached in practice) is imposed for numerical reasons.} (i.e. permitting super-Eddington accretion). Dynamical friction and BH-BH mergers are modelled following \citet{2022MNRAS.516..167B}. 

At each resolution, the \colibre{} simulations include models with purely thermal AGN feedback and hybrid AGN feedback (which combines bipolar kinetic jets with thermal energy injections and accounts for the evolution of BH spin; \citealt{2026MNRAS.547ag324H}). Thermal AGN feedback is implemented deterministically, following \citet{2009MNRAS.398...53B}, but with a modification: BH particles heat their neighbouring gas by a temperature increment proportional to the BH (subgrid) mass, $\Delta T_{\rm AGN} \propto m_{\rm BH}$, rather than by a fixed $\Delta T_{\rm AGN} $ \citep[see][for details]{2026COLIBREproject}. This work uses only the simulations with purely thermal AGN feedback\footnote{Although the \colibre{} simulations with the hybrid AGN feedback prescription model BH and AGN physics in greater detail than those with thermal AGN feedback, we do not use the hybrid model in this work because BH growth in the hybrid model appears to be too slow at high redshift, leading to insufficient quenching of massive galaxies \citep[see fig. 13 in][for details]{chaikin2026evolutiongalaxystellarmass}. The slower BH growth is likely a consequence of the fact that super-Eddington accretion in the hybrid model may be suppressed too strongly \citep[see $\S$6.2 in][for details and possible improvements]{2026MNRAS.547ag324H}.}, which is the fiducial AGN feedback model in \colibre. In the thermal model, the radiative efficiency of the subgrid accretion disc is assumed to be 10 per cent, and the BH seed mass at m7 resolution is $10^{5.5}~\mathrm{M_\odot}$. The strengths of the \colibre{} SN and AGN energy feedback were optimised to reproduce the observed $z = 0$ GSMF from GAMA DR4 \citep{2022MNRAS.513..439D}, the observed $z \approx 0$ galaxy stellar mass -- size relation reported by \citet{2022MNRAS.509.3751H}, and the $z\approx 0$ masses of BHs in massive galaxies, as detailed in \citet{2026COLIBREcalibration} for the thermal AGN feedback and in \citet{2026MNRAS.547ag324H} for the hybrid feedback simulations.

The \colibre{} simulations produce 128 data outputs between $z=30$ and $0$ \citep[see][for details]{2026COLIBREproject}. Subhaloes are identified using the history-based halo finder HBT-HERONS \citep{2018MNRAS.474..604H,2025MNRAS.543.1339F}. The HBT-HERONS output is processed with the Spherical Overdensity and Aperture Processor (SOAP; \citealt{2025JOSS...10.8252M}) to generate a comprehensive catalogue of halo and galaxy properties. From SOAP, we use galaxy stellar mass $M_*$ and star formation rate (SFR), computed by summing the masses of stellar particles and the instantaneous SFRs of gas particles, respectively. These particles must be gravitationally bound to the subhalo and located within a 3D spherical aperture of radius $50$~pkpc, which is the default choice in \colibre. For the $M_*$ -- BH mass relation, we take the (subgrid) mass of the BH particle that is bound to the subhalo and within $50$~pkpc. If a subhalo contains more than one such BH, the mass of the most massive BH is used; if no BH is bound, the BH mass assigned to that subhalo is zero. Finally, all properties intrinsic to BH particles, such as the instantaneous Eddington fraction and the cumulative energy injected by AGN feedback, are taken directly from the snapshot output.

\begin{figure*}
    \centering
    \includegraphics[width=0.999\linewidth]
    {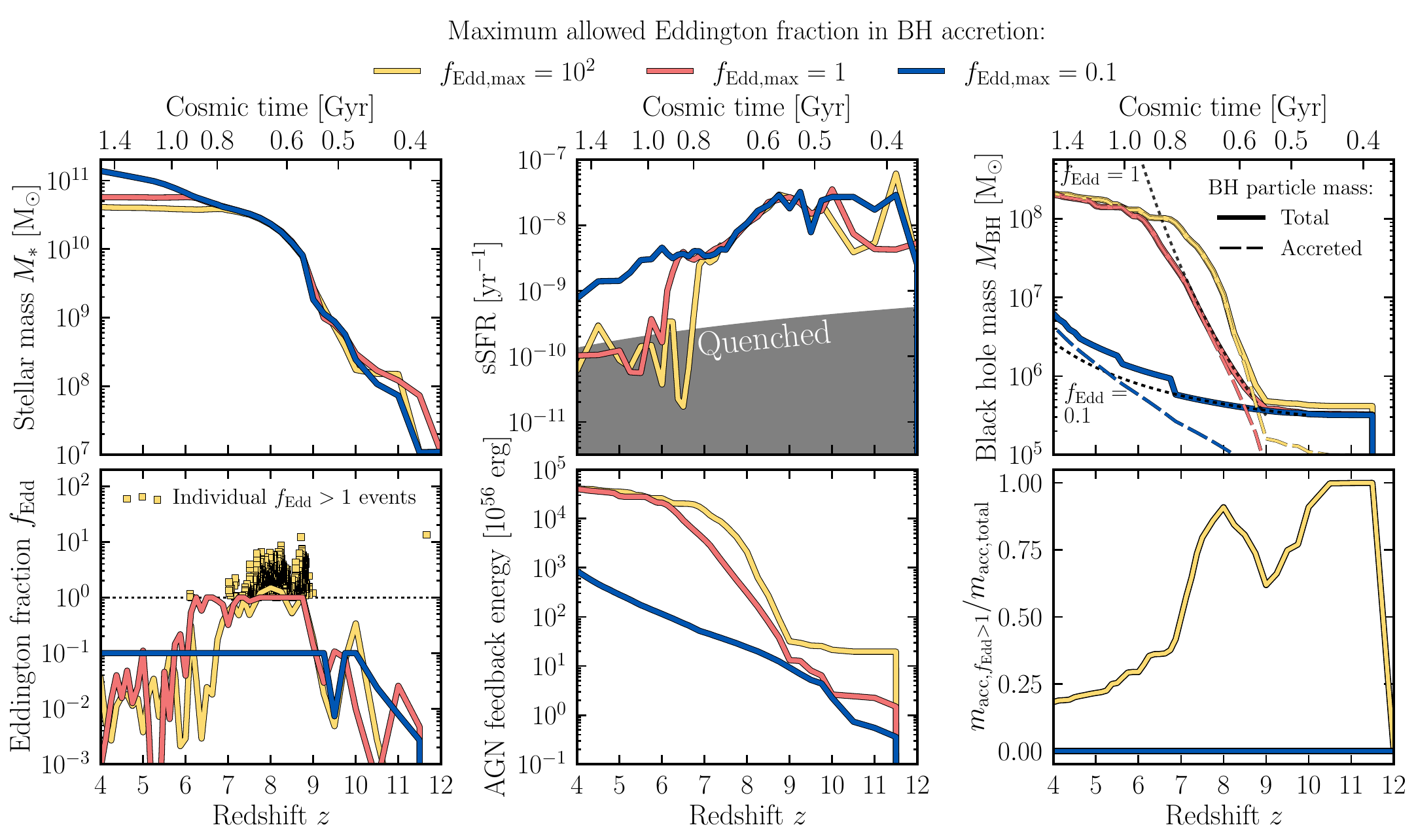}
    \vspace{-0.7cm}
    \caption{An example of a representative massive galaxy whose star formation is quenched by AGN feedback in the simulations with different maximum allowed Eddington fractions: $f_{\rm Edd,max} = 10^2$ (yellow), $1$ (red), and $0.1$ (blue). In all three cases, the evolution of the same galaxy is shown. From left to right, the top row shows the galaxy stellar mass, its sSFR, and its BH mass. The bottom row shows the Eddington fraction, the cumulative AGN feedback energy injected by the BH, and the cumulative fraction of mass accreted by the BH in $f_{\rm Edd} > 1$ accretion events (see the main text for details). Allowing super-Eddington accretion results in enhanced BH growth at high redshift, leading to the galaxy becoming quenched by AGN feedback at earlier times.}
\label{fig: representative galaxy}
\end{figure*}

\begin{figure*}
    \centering
    \includegraphics[width=0.999\linewidth]{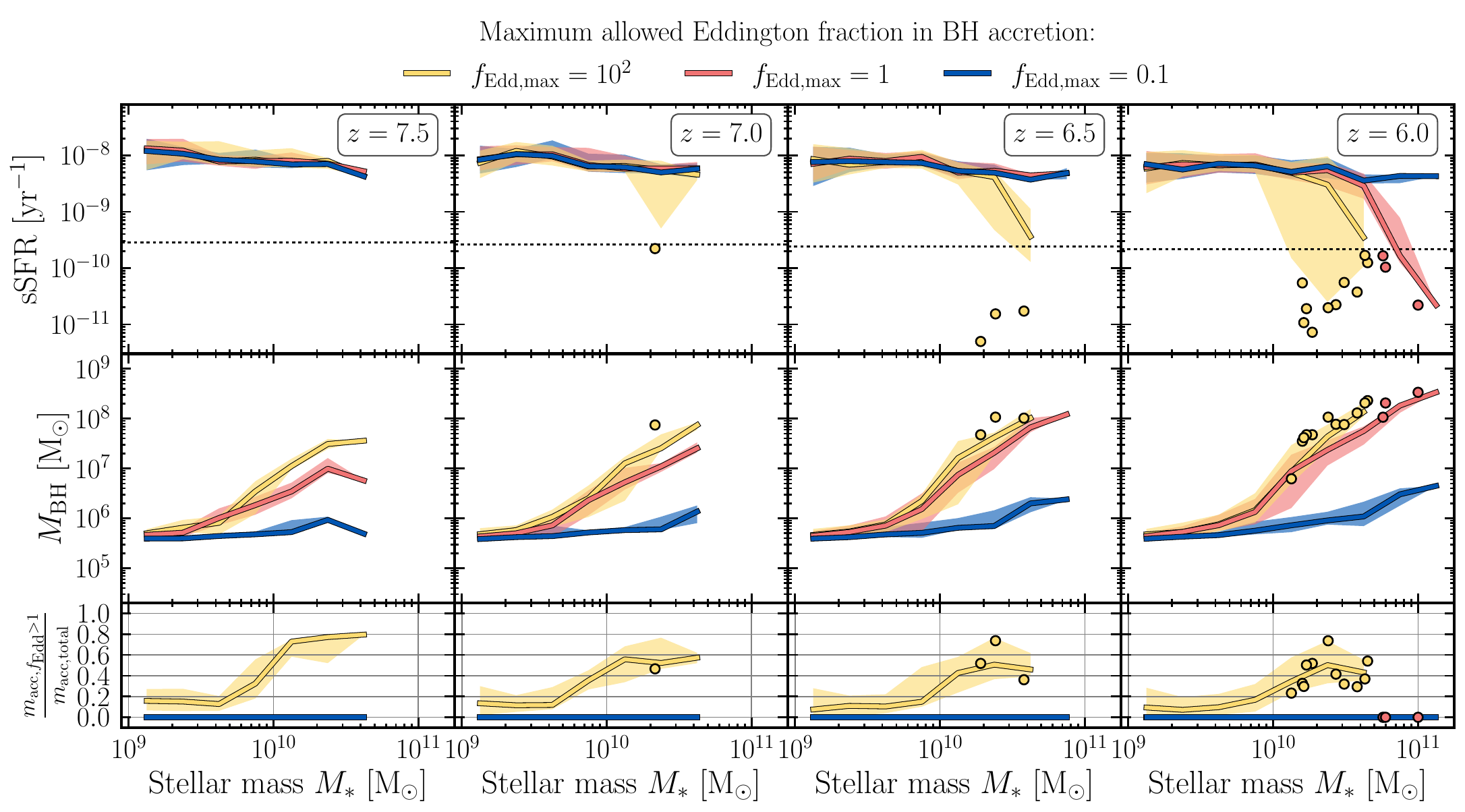}
    \vspace{-0.7cm}
    \caption{Evolution of the sSFR (top), BH mass (middle), and the cumulative fraction of BH mass accreted at $f_{\mathrm{Edd}} > 1$, all plotted against stellar mass. Results are shown for simulations with $f_{\rm Edd,max}=10^2$, $1$, and $0.1$ (colours). Columns show different redshifts (left to right): $7.5$, $7$, $6.5$, and $6$. Solid lines indicate the median relations and shaded regions the 16$^{\rm th}$ to 84$^{\rm th}$ percentiles, all computed in $0.2$-dex $M_*$ bins. In the top panels, the black horizontal lines mark the quenching threshold of $0.2/t_{\rm age}$. Individual circles denote all MQGs satisfying $\text{sSFR} < 0.2/t_{\rm age}$ and $M_* > 10^{9.5}~\mathrm{M_\odot}$. The enhanced BH growth due to super-Eddington accretion is evident across the entire population of massive galaxies, enabling their earlier quenching by AGN feedback and leading to a larger number of MQGs.}
\label{fig: evolution of sSFR and BH mass}
\end{figure*}

\begin{figure}
    \centering
    \includegraphics[width=0.99\linewidth]{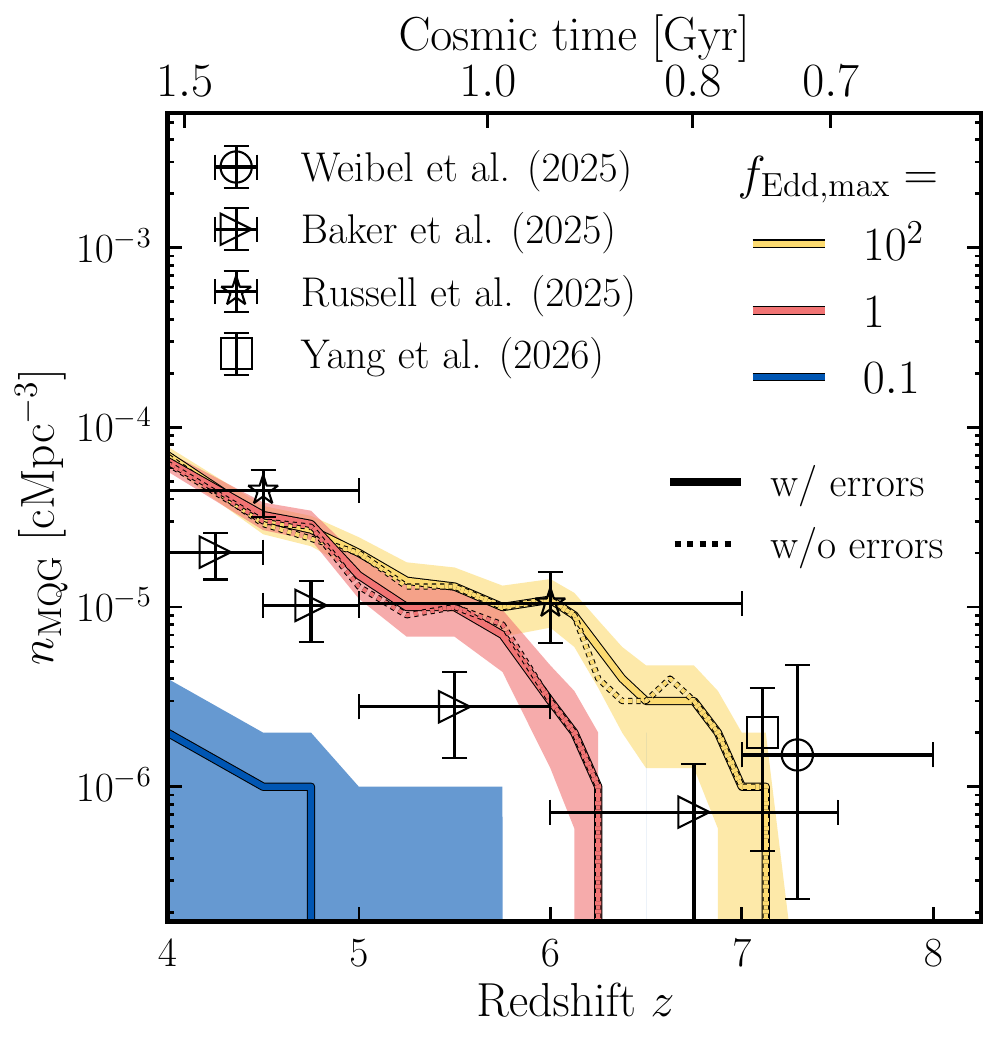}
    \vspace{-0.25cm}
    \caption{Evolution of the comoving number density of MQGs in simulations with different maximum allowed Eddington fractions: $f_{\rm Edd,max}=10^2$ (yellow), $1$ (red), and $0.1$ (blue). Solid curves show predictions after adding $0.3$~dex lognormal errors to SFRs and $M_*$; dotted curves show results without errors. Black symbols indicate \textit{JWST} constraints. Only the fiducial model ($f_{\rm Edd,max}=10^2$), which allows super-Eddington BH accretion, is consistent with the \textit{JWST} data at $z \gtrsim 6$.}
    \label{fig:n_MQG_vs_data}
\end{figure}

\begin{figure}
    \centering
    \includegraphics[width=0.99\linewidth]{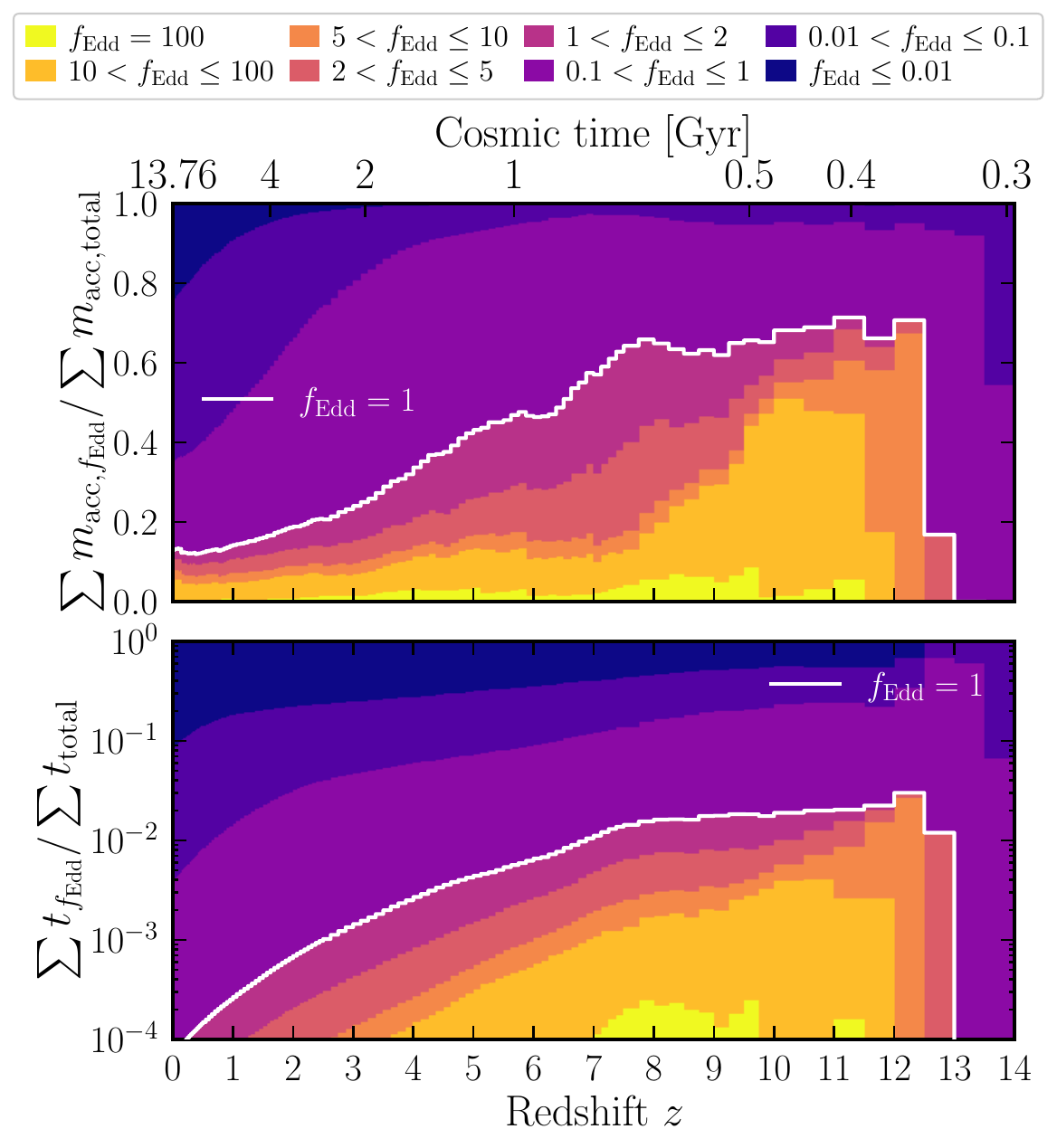}
    \vspace{-0.25cm}
    \caption{Cumulative fraction of BH mass accreted (top panel) and time spent (bottom panel) in accretion events at different Eddington fractions (colour-coded). Results are shown for the model with $f_{\rm Edd,max} = 10^2$, using all BH particles in the simulation to calculate the cumulative fractions. The white curve marks $f_{\rm Edd} = 1$ for reference. Although super-Eddington accretion ($f_{\rm Edd} > 1$) contributes more than $\approx 50$ per cent of the BH mass growth at high redshift, with $f_{\rm Edd} \gtrsim 10$ frequently reached, BHs spend only a few per cent of their time in the super-Eddington regime.}
    \label{fig: accr mass and time spent vs z}
\end{figure}

\subsection{Model variations}

In \colibre, the BH accretion rate, $\dot{m}_{\rm accr}$, is capped at $100$ times the Eddington rate, $\dot{m}_{\rm Edd}$ (i.e. the Eddington fraction $f_{\rm Edd} \equiv \dot{m}_{\rm accr}/\dot{m}_{\rm Edd}$ cannot exceed $f_{\rm Edd,max} = 10^2$), with $\dot{m}_{\rm accr}$ and $\dot{m}_{\rm Edd}$ defined by equations (23) and (26) in \citet{2026COLIBREproject}, respectively. To demonstrate the importance of super-Eddington accretion, we ran two new simulations where (i) we capped the accretion rate at the Eddington limit ($f_{\rm Edd,max}=1$) and (ii) at 10 per cent thereof ($f_{\rm Edd,max}=0.1$). These two models were run in a $100^3~\textrm{cMpc}^3$ volume at m7 resolution, with all other parameters matching the fiducial \colibre{} m7 model with thermal AGN feedback, for which $f_{\rm Edd,max}=10^2$. We chose this volume and resolution as it was our only computationally feasible option that could produce realistic non-zero number densities of MQGs at high $z$ ($\sim 10^{-6}~\mathrm{cMpc}^{-3}$). For consistency with the variation simulations, we use the fiducial simulation with the same volume and resolution\footnote{More precisely, instead of using the L100m7 simulation from the original \colibre{} suite, we re-ran this simulation with extra BH-related output, which we needed in this work. We verified that the results from the two simulations are statistically indistinguishable.}, although the fiducial model is also available in $8\times$ and $64\times$ larger volumes \citep[see table 2 in][]{2026COLIBREproject}. 

\section{Results}
\label{section: results}

Fig.~\ref{fig: representative galaxy} shows the evolution of several properties of a representative massive galaxy and its central BH from $z=12$ to $4$. Each panel shows three solid curves, corresponding to our fiducial model with $f_{\rm Edd,max} = 10^2$ and its variations with $f_{\rm Edd,max} = 1$ and $0.1$. The same galaxy, cross-matched across the three L100m7 simulations, which use identical initial conditions, is shown in all cases. The galaxy resides in a central halo whose total mass grows from $M_{\rm 200c}\sim 10^{10}~\mathrm{M_\odot}$ at $z = 12$ to $\sim 10^{12.5}~\mathrm{M_\odot}$ by $z = 4$. The top left panel shows the galaxy stellar mass $M_*$, the top middle panel its instantaneous sSFR ($\equiv \text{SFR}/M_*$), and the top right panel the mass of its BH. In addition, the three long-dashed curves in the top right panel indicate the accreted BH mass in the simulations (i.e. excluding the seed mass and mass gained via mergers), while the two black dotted lines show the theoretical BH mass evolution for continuous accretion at $f_{\rm Edd}=0.1$ and $1$ (without mergers), starting at $z=10$ and $z=9$, respectively.

In the three simulations, the galaxy and its BH evolve nearly identically from $z=12$ until $z\approx 9$, with minor differences resulting from stochastic effects \citep[e.g.][]{2023MNRAS.526.2441B}. Below $z \approx 9$, the solid curves in the top right panel start to diverge as the BHs begin to grow rapidly via gas accretion, with the models with lower $f_{\rm Edd,max}$ predicting slower BH growth. Comparison of the yellow, red, and blue curves to the black dotted lines between $z\approx 7$ and $9$ confirms that the BH in the fiducial model undergoes super-Eddington growth, while the growth in the $f_{\rm Edd,max}=1$ and $f_{\rm Edd,max}=0.1$ models proceeds at $\approx 100$ and $10$ per cent of the Eddington rate. The faster BH growth triggers an earlier onset of AGN feedback. Consequently, the galaxy in the $f_{\rm Edd,max}=10^2$ simulation becomes quiescent below $z \approx 7$. In contrast, with $f_{\rm Edd,max}=1$, the same galaxy continues actively forming stars until $z \approx 6$, while with $f_{\rm Edd,max}=0.1$, it remains star-forming throughout the entire redshift range explored. 

The bottom row of Fig.~\ref{fig: representative galaxy} provides further insight into the properties of the BH: the instantaneous Eddington fraction $f_{\rm Edd}$ (left); the cumulative AGN feedback energy injected into the nearby gas (middle); and the cumulative fraction of mass accreted in the super-Eddington regime (right), which is defined for each redshift as the mass accreted at $f_{\rm Edd} > 1$ up to that redshift, $m_{\mathrm{acc}, f_{\rm Edd}>1}$, divided by the total accreted mass, $m_{\rm acc, total}$. The yellow squares in the left panel mark individual super-Eddington accretion events (i.e. time-steps with $f_{\rm Edd} > 1$) experienced by the BH in the fiducial simulation, which have a much higher cadence than the standard \colibre{} output. The BH in the fiducial model experiences numerous super-Eddington accretion events between $z=9$ and $7$, with the highest rate exceeding the Eddington limit by a factor $\sim 10$. In the two variation models, $f_{\rm Edd}$ hits a ceiling at the corresponding $f_{\rm Edd,max}$ value around $z\approx 9$ and follows it until $z\approx 6$ ($z = 4$) for $f_{\rm Edd,max}=1$ ($f_{\rm Edd,max}=0.1$).

Once the BH has injected enough AGN energy to quench the star formation in its host galaxy, $f_{\rm Edd}$ drops to $\sim 10^{-3} - 10^{-1}$, which occurs below $z\approx 7$ and $z\approx 6$ in the $f_{\rm Edd,max}=10^2$ and $f_{\rm Edd,max}=1$ models, respectively. After this point, the rapid gas accretion phase halts, and the BH subsequently grows slowly in a self-regulated regime. Although, for this particular galaxy, the $f_{\rm Edd,max}=1$ model predicts an initially slower BH growth compared to $f_{\rm Edd,max}=10^2$, the earlier quenching for $f_{\rm Edd,max}=10^2$ allows the BH in the $f_{\rm Edd,max}=1$ model to catch up in mass, resulting in nearly identical final BH masses at $z\lesssim 6$. Finally, the bottom right panel shows that, at the time when the BH quenches its host galaxy in the $f_{\rm Edd,max}=10^2$ model, about 50 per cent of its accreted mass was accreted at super-Eddington rates.

The results from Fig.~\ref{fig: representative galaxy} for $f_{\rm Edd,max}=10^2$ are generally in line with findings from higher-resolution cosmological zoom-in simulations of massive galaxies in which the central BH is allowed to grow at super-Eddington rates. For example, \citet{2025A&A...704A.248Q} simulated the evolution of a massive ($M_{\rm halo}\approx 3\times10^{12}~\mathrm{M_\odot}$ at $z=6$) halo down to $z=7.5$ with a gas mass resolution in the zoom-in region of $m_{\rm gas}\sim 10^{4}~\mathrm{M_\odot}$, finding that the BH undergoes super-Eddington accretion at $8.6 \lesssim z \lesssim 9.5$, during which it grows by more than two orders of magnitude in mass. By the end of this phase, powerful AGN feedback carves out a cavity within the central 200 pc, thereby quenching the galaxy until new gas flows in from larger scales. Similarly, \citet{2025MNRAS.537.2559H} simulated the evolution of a massive halo down to $z=4.3$ (where $M_{\rm halo}\approx 10^{13}~\mathrm{M_\odot}$) with a gas particle mass in the zoom-in region of $m_{\rm gas}\sim 10^{6}~\mathrm{M_\odot}$, finding that the central BH undergoes nearly continuous super-Eddington accretion at $7.5 < z < 8.5$, leading by $z=7.5$ to an increase in its mass by more than two orders of magnitude and temporary quenching of the host galaxy.

Having illustrated the main differences between models with different $f_{\rm Edd,max}$ for a single massive galaxy, we next examine the evolution of all massive galaxies in the simulated volumes. Fig.~\ref{fig: evolution of sSFR and BH mass} shows the redshift evolution from $z=7.5$ to $6$ of the sSFR -- $M_*$ relation (top row); the BH mass -- $M_*$ relation (middle row); and the cumulative fraction of BH mass accreted at $f_{\rm Edd} > 1$, $m_{\mathrm{acc}, f_{\rm Edd}>1}/m_{\rm acc, total}$, versus $M_*$ (bottom row). At $z=7.5$, the sSFR -- $M_*$ relation is indistinguishable between the three models, with all galaxies following the star-forming main sequence (SFMS). Clear differences emerge, however, in the BH mass -- $M_*$ relation: the masses of BHs in massive galaxies ($M_* \gtrsim 10^{10}~\mathrm{M_\odot}$) are $\approx 0.5$–$0.7$~dex higher for $f_{\rm Edd,max}=10^2$ than for $f_{\rm Edd,max}=1$, while in the $f_{\rm Edd,max}=0.1$ model, all BHs remain close to their seed mass of $10^{5.5}~\mathrm{M_\odot}$. At $z=7$, the $f_{\rm Edd,max}=10^2$ model produces its first MQG (yellow circle), while galaxies in the other two models remain on the SFMS. By $z=6.5$, the fiducial model predicts three MQGs, whereas the others continue to follow the SFMS. Only at $z=6$ does the $f_{\rm Edd,max}=1$ model produce its first MQGs, with a number a factor of three lower than in the fiducial model. In both models, the BH masses in most MQGs are higher than the median BH mass -- $M_*$ relation, indicating the dominant role of AGN feedback in driving the quenching.

Fig.~\ref{fig:n_MQG_vs_data} compares the predicted number density of MQGs in the simulations, $n_{\rm MQG}$, with recent \textit{JWST} constraints at $z > 6$. We include data from \citet{2025A&A...702A.270B} and \citet{2025MNRAS.544.4482R}, who define MQGs as $M_* > 10^{9.5}~\mathrm{M_\odot}$ and $\mathrm{sSFR} < 0.2/t_{\rm age}$; from \citet{2026ApJ..1000L..42Y}, corresponding to $10^{9.5} < M_*/\mathrm{M_\odot} < 10^{10.6}$ and $\mathrm{sSFR} < 0.2/t_{\rm age}$; and a single galaxy-based estimate from \citet{2025ApJ...983...11W}, with $M_* \approx 10^{10.2}~\mathrm{M_\odot}$ and $\mathrm{sSFR} < 10^{-10}~\mathrm{yr}^{-1}$. To maximize consistency with these selection criteria, we define MQGs in the simulations as galaxies with $M_* > 10^{9.5}~\mathrm{M_\odot}$ and $\mathrm{sSFR} < 0.2/t_{\rm age}$. To ensure a consistent comparison with \citet{2026ApJ..1000L..42Y} whose selection excludes galaxies with $M_* > 10^{10.6}~\mathrm{M_\odot}$, we show their data only at $z>6.5$, where our and their criteria effectively align because at $z>6.5$ all simulated MQGs have stellar masses $10^{10} < M_*/\mathrm{M_\odot} < 10^{10.6}$ (see Fig.~\ref{fig: evolution of sSFR and BH mass}). To account for measurement uncertainties present in observations, we add errors to simulated SFRs and $M_*$ sampled from a lognormal distribution with zero mean and a standard deviation of $0.3$~dex \citep[e.g.][]{2026ApJ...997..252Z}. We repeat this process $300$ times, yielding a distribution of $n_{\rm MQG}$ values in each redshift bin. We then plot the median values (solid curves), the maximum of the 16$^{\rm th}$–84$^{\rm th}$ percentile range and Poisson uncertainties (shaded regions), and the predictions without measurement uncertainties (dotted curves). We note that the added uncertainties have a significant effect on $n_{\rm MQG}$ only for the $f_{\rm Edd,max}=0.1$ model.

Only the fiducial model, which allows super-Eddington accretion, is consistent with the \textit{JWST} data at $z>6$, in line with the earlier results from the \colibre{} simulations in larger volumes \citep{chaikin2026evolutiongalaxystellarmass,2025arXiv251216208C}. In contrast, the $f_{\rm Edd,max}=1$ model predicts no MQGs at $z > 6.25$, while $f_{\rm Edd,max}=0.1$ struggles to produce even a single MQG (none are present if no measurement errors are included). The $f_{\rm Edd,max}=10^2$ and $f_{\rm Edd,max}=1$ models converge below $z\approx 5$, as BHs in the $f_{\rm Edd,max}=1$ model eventually catch up in mass (and therefore in the energy injected by AGN feedback) with those in the fiducial model (as shown in Fig.~\ref{fig: representative galaxy}). At $z<5$, the converged models are consistent with \citet{2025MNRAS.544.4482R} but lie $\approx 0.5$~dex above \citet{2025A&A...702A.270B} \citep[see][for details]{chaikin2026evolutiongalaxystellarmass,2025arXiv251216208C}. At $z=0$, the GSMF, galaxy size -- $M_*$ relation, and BH mass -- $M_*$ relation -- the three relations used to calibrate SN and AGN feedback in \colibre{} -- are statistically indistinguishable between the $f_{\rm Edd,max}=10^2$ and $f_{\rm Edd,max}=1$ models (not shown).

Finally, the top (bottom) panel of Fig.~\ref{fig: accr mass and time spent vs z} shows the cumulative fraction of BH mass accreted (time spent) in accretion events with different Eddington fractions for the fiducial model. The numerator and denominator of the cumulative fractions are computed as sums over all BH particles in the simulation. The fraction of time spent by BHs accreting at $f_{\rm Edd}>1$ is at most a few per cent\footnote{Restricting the calculation of the cumulative fractions to only BHs with masses $\geq 10^{6}~\mathrm{M_\odot}$ would increase the cumulative fraction of time spent in super-Eddington accretion events by up to a factor of $\approx 8$ at high redshift, reaching $\approx 6$ ($\approx 10$)~per cent at $z\approx6$ ($z\gtrsim 7.5$), while having only a minor impact on the cumulative fraction of mass accreted at $f_{\rm Edd} > 1$ at all redshifts.} at high redshift and decreases to negligible values by $z=0$, even though these super-Eddington events account for a significant fraction of BH mass growth ($50-70$ per cent at $z \gtrsim 7$, as can be seen in this figure and Figs.  \ref{fig: representative galaxy} and \ref{fig: evolution of sSFR and BH mass}). In other words, this figure shows that although super-Eddington accretion is not sustained over long time-scales, it is responsible for substantial BH growth at high redshift. 

Interestingly, the fraction of time spent in $f_{\rm Edd}>1$ accretion events found in Fig.~\ref{fig: accr mass and time spent vs z} is comparable to quasar duty cycles at $z\gtrsim 6$, which are inferred to be of the order of 1 per cent based on recent measurements of the high-redshift two-point cross-correlation function of galaxies and quasars \citep[e.g.][]{2024ApJ...974..275E,2024MNRAS.534.3155P,2026A&A...708A.160S}. At the same time, BH masses as high as $\sim 10^{8}-10^{9}~\mathrm{M_\odot}$  out to $z\approx 7-9$ are inferred from measured emission line widths or X-ray flux \citep[e.g.][]{2023ApJ...957L...7K,2024NatAs...8..126B}, requiring continuous near-Eddington accretion for BHs with a seed mass of $m_{\rm seed} \sim 10^{4}~\mathrm{M_\odot}$ to reach the inferred masses by the observed redshift. A natural solution that satisfies these two sets of observational constraints is therefore super-Eddington growth during short periods of time at high redshift, in line with our Fig.~\ref{fig: accr mass and time spent vs z}, although higher obscured quasar fractions \citep[e.g.][]{2023MNRAS.521.3108S} or higher initial BH seed masses \citep[see][for a review]{2020ARA&A..58...27I} can also provide a solution.

\section{Conclusions}
\label{section: conclusions}

We have investigated the importance of super-Eddington BH growth for alleviating the tension between galaxy formation models and $z\gtrsim6$ \textit{JWST} constraints on the abundance of MQGs. Using the \colibre{} galaxy formation model \citep{2026COLIBREproject,2026COLIBREcalibration}, we ran two simulations in a $(100~\mathrm{cMpc})^{3}$ volume with BH accretion rates capped at Eddington fraction of $f_{\rm Edd,max}=1$ and $0.1$, complementing the fiducial model where the cap is $f_{\rm Edd,max}=10^{2}$ (i.e. allowing super-Eddington accretion). Our main findings are as follows:

\begin{itemize}
 
\item Super-Eddington accretion is readily realized in the fiducial \colibre{} model during the rapid phase of gas accretion onto BHs at high redshift, with Eddington fractions of $f_{\rm Edd} \sim 10$ frequently reached (Fig.~\ref{fig: representative galaxy}). This significantly enhances BH mass growth, leading to an earlier onset of AGN feedback and, consequently, to earlier quenching of the massive galaxies hosting the BHs.

\item The enhanced BH growth at high redshift due to super-Eddington accretion is evident across the entire population of massive galaxies (Fig.~\ref{fig: evolution of sSFR and BH mass}). By $z = 6$, as much as $\approx 50$ per cent of the accreted mass of BHs in massive galaxies has been accreted at super-Eddington rates, even though BHs spend only a few per cent of their time accreting at $f_{\rm Edd} > 1$ (Fig.~\ref{fig: accr mass and time spent vs z}).

\item The enhanced BH growth renders the fiducial \colibre{} model consistent with \textit{JWST} constraints on the number density of MQGs at $z \gtrsim 6$, whereas the $f_{\rm Edd,max}=1$ and $f_{\rm Edd,max}=0.1$ models strongly undershoot the high-redshift data (Fig.~\ref{fig:n_MQG_vs_data}).
\end{itemize}

Taken together, the results of our cosmological simulations provide strong evidence that super-Eddington accretion offers a promising solution to the tension between predictions from galaxy formation simulations adopting the $\Lambda$CDM cosmology and the high-redshift number densities of MQGs revealed by \textit{JWST}.

In closing, we emphasize that the \colibre{} simulations reproduce the observed evolution of the GSMF at $0 < z < 12$ \citep{chaikin2026evolutiongalaxystellarmass} and are consistent with the sizes and star formation histories of observed MQGs at high redshift \citep{2025arXiv251216208C}. This agreement suggests that, although BH physics in \colibre{} is necessarily captured rather crudely given the limited simulation resolution, the effective impact of AGN feedback on resolved galactic scales is not unrealistic. The analysis in this work is subject to cosmic variance, as the simulated volume is relatively small ($100^{3}~\mathrm{cMpc}^{3}$). However, \citet{chaikin2026evolutiongalaxystellarmass} show that the fiducial \colibre{} model remains consistent with the $z>6$ \textit{JWST} data in $8\times$ and $64\times$ larger volumes. 

Finally, we note that in \colibre, super-Eddington accretion is not limited to m7 resolution, but is also an important mechanism for BH growth at the higher m6 and m5 resolutions. However, even though the largest \colibre{} simulations at m7 and m6 resolution are both consistent with the observed abundances of MQGs at high $z$, the quenching of massive galaxies in \colibre{} becomes systematically weaker at higher resolutions (see figs.~9 and 11 in \citealt{chaikin2026evolutiongalaxystellarmass} and the accompanying text for discussion of the lack of good convergence), and there are hints that the simulations at m5 resolution, which are currently not available in sufficiently large volumes, could undershoot the data. This may suggest that allowing super-Eddington BH growth is a necessary, but not sufficient, condition for reproducing the observed high-$z$ abundances of MQGs.

\section*{Acknowledgements}

We thank the anonymous referee for their constructive feedback. This work used the DiRAC@Durham facility managed by the Institute for Computational Cosmology on behalf of the STFC DiRAC HPC Facility (www.dirac.ac.uk). The equipment was funded by BEIS capital funding via STFC capital grants ST/K00042X/1, ST/P002293/1, ST/R002371/1 and ST/S002502/1, Durham University and STFC operations grant ST/R000832/1. DiRAC is part of the National e-Infrastructure. This project has received funding from the Netherlands Organization for Scientific Research (NWO) through research programme Athena 184.034.002. EC acknowledges support from STFC consolidated grant ST/X001075/1. CGL acknowledges support from STFC consolidated grant ST/X001075/1. SP acknowledges support by the Austrian Science Fund (FWF) through grant-DOI: 10.55776/V982. The research in this paper made use of the \textsc{Swift} open-source simulation code (\url{http://www.swiftsim.com}, \citealt{2024MNRAS.530.2378S}) version 1.0.0.

\section*{Data Availability}

The data underlying this article will be shared on reasonable request to the corresponding author. The public version of the \textsc{Swift} simulation code can be found on \href{http://www.swiftsim.com}{www.swiftsim.com}. The \textsc{Swift} modules related to the \colibre{} galaxy formation model will be integrated into the public version after the public release of \colibre. The \textsc{chimes} astrochemistry code is publicly available at \href{https://richings.bitbucket.io/chimes/home.html}{https://richings.bitbucket.io/chimes/home.html}. The HBT-HERONS  halo finder is available at \url{https://hbt-herons.strw.leidenuniv.nl/}.

\bibliographystyle{mnras}
\bibliography{main} 

\appendix

% Don't change these lines
\bsp	% typesetting comment
\label{lastpage}
\end{document}